\pdfoutput=1

\long\def\symbolfootnote[#1]#2{\begingroup%
\def\thefootnote{\fnsymbol{footnote}}\footnote[#1]{#2}\endgroup}

\documentclass[12pt]{article}
\usepackage[note,natbib]{achemso}  
\usepackage{graphicx} 
\usepackage{dcolumn}
\usepackage{bm}
\usepackage{amsmath,amsfonts,amssymb,afterpage}
\usepackage{hyperref}  
\usepackage[figure]{hypcap}  

\begin{document}

\title{\huge{Alignment Dynamics of Single-Walled Carbon Nanotubes in Pulsed Ultrahigh Magnetic Fields}}
\maketitle

\begin{centering}

\author{
	\textit{
		Jonah~Shaver~$^{1,\dagger}$,
		A.~Nicholas~G.~Parra-Vasquez~$^{2,\dagger}$,
 		Stefan~Hansel~$^{3,4}$,
 		Oliver~Portugall~$^4$,
 		Charles~H.~Mielke~$^5$,
 		Michael~von~Ortenberg~$^3$,
 		Robert~H.~Hauge~$^6$,
		Matteo~Pasquali~$^2$,
 		and Junichiro~Kono~$^{1,*}$
 	}
 }

~\\
\noindent$^1$Department of Electrical and Computer Engineering, Rice University, Houston, Texas 77005;
$^2$Department of Chemical and Biomolecular Engineering, Rice University, Houston, Texas 77005;
$^3$Institut f\"ur Physik, Humboldt-Universit\"at zu Berlin, Berlin, Germany;
$^4$Laboratoire National des Champs Magn\'{e}tiques Puls\'{e}s, 31400 Toulouse, France;
$^5$National High Magnetic Field Laboratory, Los Alamos, New Mexico 87545;
$^6$Department of Chemistry, Rice University, Houston, Texas 77005
~\\
\date{\textbf{\today}}

\end{centering}

~\\
\noindent$^\dagger$Jonah~Shaver and A.~Nicholas~G.~Parra-Vasquez contributed equally to this work.
$^*$Please address all correspondence to kono@rice.edu.
\\


\newpage

\begin{abstract}
We have measured the dynamic alignment properties of single-walled carbon nanotube (SWNT) suspensions in pulsed high magnetic fields through linear dichroism spectroscopy.  Millisecond-duration pulsed high magnetic fields up to 56~T as well as microsecond-duration pulsed ultrahigh magnetic fields up to 166~T were used.  Due to their anisotropic magnetic properties, SWNTs align in an applied magnetic field, and because of their anisotropic optical properties, aligned SWNTs show linear dichroism.  The characteristics of their overall alignment depend on several factors, including the viscosity and temperature of the suspending solvent, the degree of anisotropy of nanotube magnetic susceptibilities, the nanotube length distribution, the degree of nanotube bundling, and the strength and duration of the applied magnetic field.  In order to explain our data, we have developed a theoretical model based on the Smoluchowski equation for rigid rods that accurately reproduces the salient features of the experimental data.
\end{abstract}

\textbf{Single-walled carbon nanotubes (SWNTs), rolled up tubes of graphene sheets, are unique nano-objects with extreme aspect ratios, which lead to unusually anisotropic electrical, magnetic, and optical properties.
They can be individually suspended in aqueous solutions with appropriate surfactants,\cite{OconnelletAl02Science} and such suspended SWNTs behave roughly as rigid rods undergoing Brownian motion\cite{duggal:246104}.  In the absence of external fields, their orientation angles are randomly distributed.  However, when placed in a perturbing field, suspended SWNTs will align parallel to the field lines due to their anisotropic properties.  The steady state alignment of SWNTs in magnetic,\cite{FujiwaraetAl01JPCA,ZaricetAl04Science,ZaricetAl04NL,IslametAl04PRL,islam:201401,TorrensO.N._ja066719+} electric,\cite{fagan:213105} flow,\cite{Davis2004,Hobbie04JCP,ParraVasquez:2007p162,CaseyJohnP._nn800351n} and strain fields\cite{faganetal07prl} has been characterized in many recent studies.  Though mentions of dynamic alignment have been made,\cite{ZaricetAl06PRL,Shaver:2005p87} to date there are no comprehensive studies. Here we present the first combined experimental and theoretical study that provides fundamental insight into the hydrodynamic motion of these highly-anisotropic nano-objects.}

The magnetic susceptibilities of SWNTs of different diameters, chiralities, and types have been theoretically calculated using different methods.\cite{AjikiAndo93BJPSJ,Lu95PRL,AjikiAndo95JPSJ,MarquesetAl06PRB}  Semiconducting SWNTs are predicted to be diamagnetic ($\chi<0$) both parallel ($\parallel$) and perpendicular ($\perp$) to their long axis, but the perpendicular susceptibility is predicted to have a larger magnitude ($|\chi_\perp|>|\chi_\parallel|$), aligning the SWNT parallel to the field.  Metallic SWNTs are predicted to be paramagnetic (diamagnetic) parallel (perpendicular) to their long axes ($\chi_\parallel > 0$, $\chi_\perp < 0$) and thus also align parallel to the applied field.  For $\sim$1-nm-diameter nanotubes, the values for the magnetic anisotropy, $\Delta\chi = \chi_\perp - \chi_\parallel$, calculated by an {\em ab initio} method are between 1.2 and 1.8 $\times$ 10$^{-5}$~emu/mol, depending on the tube chirality,\cite{MarquesetAl06PRB} which are similar to the values calculated by a ${\bf k}\cdot {\bf p}$ method (1.9 $\times$ 10$^{-5}$~emu/mol)\cite{AjikiAndo95JPSJ} and by a tight-binding method (1.5 $\times$ 10$^{-5}$~emu/mol).\cite{Lu95PRL}  These values are consistent with recently-reported experimental values, measured with steady-state optical methods.\cite{ZaricetAl04NL,TorrensO.N._ja066719+}

The degree of alignment of SWNTs in a magnetic field can be conveniently characterized by the dimensionless ratio of the alignment potential energy and the thermal energy,
\begin{equation}
\xi=\sqrt{\frac{{B^{2}N\Delta\chi}}{k_{\rm{B}}T}}
\label{align_E}
\end{equation}
where $B$ is the magnetic field, $N$ is the number of carbon atoms in the SWNT, $k_{\rm B}$ is the Boltzmann constant, and $T$ is the temperature of the solution.  A significant fraction of nanotubes in the solution will align with $B$ when the alignment energy is greater than the randomizing energy, i.e., when $\xi>1$.  Using $u$ and the angle ($\theta$) between a SWNT and the aligning magnetic field, an angular distribution function,\cite{WaltersetAl01CPL} $P(\theta)$, in thermal equilibrium can be calculated as
\begin{equation}
\frac{dP(\theta)}{d\theta}=\frac{e^{-\xi^{2}\sin^{2}\theta}\sin\theta}{\int^{\pi/2}_{0}e^{-\xi^{2}\sin^{2}\theta}\sin\theta
d\theta}\rm{.}
\label{prob_dist}
\end{equation}
Many experiments have studied the equilibrium alignment of SWNTs in magnetic fields.\cite{TsuietAl00APL,WaltersetAl01CPL,FujiwaraetAl01JPCA,IslametAl04PRL}  More recent experiments have explored the chirality dependence of SWNT alignment to extract the SWNT species specific magnetic susceptibilities.\cite{TorrensO.N._ja066719+}

Linear dichroism spectroscopy has a well-developed history of application to both steady state and dynamic situations, such as the flow-induced alignment of fibrils\cite{Adachi:2007p86} and the magnetic-field-induced alignment of polyethylene and carbon fibers.\cite{Kimura:2000p720}  However, to date no one has studied the dynamic effects of alignment of SWNTs.  Defined as the difference between the absorbance of light polarized parallel ($A_\parallel$) and perpendicular ($A_\perp$) to the orientational director of a system, $\hat{n}$, linear dichroism ($LD$) is a measure of the degree of alignment of any solution of anisotropic molecules.\cite{dichroism}  Experimentally, the sign of $LD$ gives qualitative information about the relative orientation of molecules, positive for alignment parallel to $\hat{n}$ and negative for perpendicular.  Reduced $LD$, $LD^{\rm r}$, is normalized by the unpolarized, isotropic absorbance ($A$) of the system, and gives a quantitative measure of the alignment.  The measured $LD^{\rm r}$ spectrum is related to both the polarization of the transition moment being probed and the overall degree of alignment of the molecules being investigated:\cite{dichroism}
\begin{equation}
LD^r = \frac{LD}{A}=\frac{A_\parallel-A_\perp}{A}=3\left(\frac{3\cos^2{\alpha} - 1}{2}\right)S
\label{LDr}
\end{equation}
where $\alpha$ is the angle between the transition moment and the long axis of the molecule and $S$ is the nematic order parameter.  $S$ is a dimensionless quantity that scales from 0 for an isotropic sample to 1 for a perfectly aligned sample and is defined as
\begin{equation}
S=\frac{3\left<\cos^2{\theta}\right> - 1}{2}
\label{orderparam}
\end{equation}
where $\left<\cos^2{\theta}\right>$ is averaged over the angular probability distribution function and $\theta$ is the microscopic angle made between a SWNT's long axis and the alignment director of the system.

For the case of SWNTs, optical selection rules\cite{AjikiAndo93JPSJ} coupled with a strong depolarization for light polarized perpendicular to the tube axis result in appreciable absorption features observed only when light is polarized parallel to the tube axis.  Hence, we can simplify Eq.~(\ref{LDr}) using $\alpha=0$, to $LD^r=3S$, giving a direct link between the measured $LD^r$ and the orientation of the SWNTs.

In this study, the {\em dynamic} effects of SWNT alignment in pulsed high magnetic fields were investigated for the first time.  We measured time-dependent transmittance through individually-suspended SWNTs in aqueous solutions in the Voigt geometry (light propagation perpendicular to the applied magnetic field) in two polarization configurations, parallel and perpendicular to the applied magnetic field.  From this we calculated $LD$ as a function of time, both in millisecond (ms)-long pulsed high magnetic fields up to 56~T and microsecond ($\mu$s)-long pulsed ultrahigh magnetic fields up to 166~T.  We developed a theoretical model based on the Smoluchowski equation, which extracts the length distribution of the SWNTs in suspension based on a fit to time-dependent $LD$.  These results pave the way to further study of SWNT dynamics in solution.

\section*{Results}
\subsection*{Measured Transmittance}
All ms-pulse data was taken using a spectrally resolved, near-infrared setup.  To avoid any convolution with spectral lineshape broadening and splitting~\cite{ZaricetAl04Science,ZaricetAl04NL,ZaricetAl06PRL,Shaver:2008p215} the data was integrated over the entire InGaAs range ($\sim$900~nm to 1800~nm).   The benefit of removing ambiguity associated with spectral changes induced by the Aharonov-Bohm~\cite{Ando06JPSJ} effect coupled with the large number of nanotube chiralities present in our sample outweighs the possibility for any chirality selective analysis (which has been performed at low magnetic fields~\cite{TorrensO.N._ja066719+}).
\begin{figure}[!h]
\includegraphics[scale=.7]{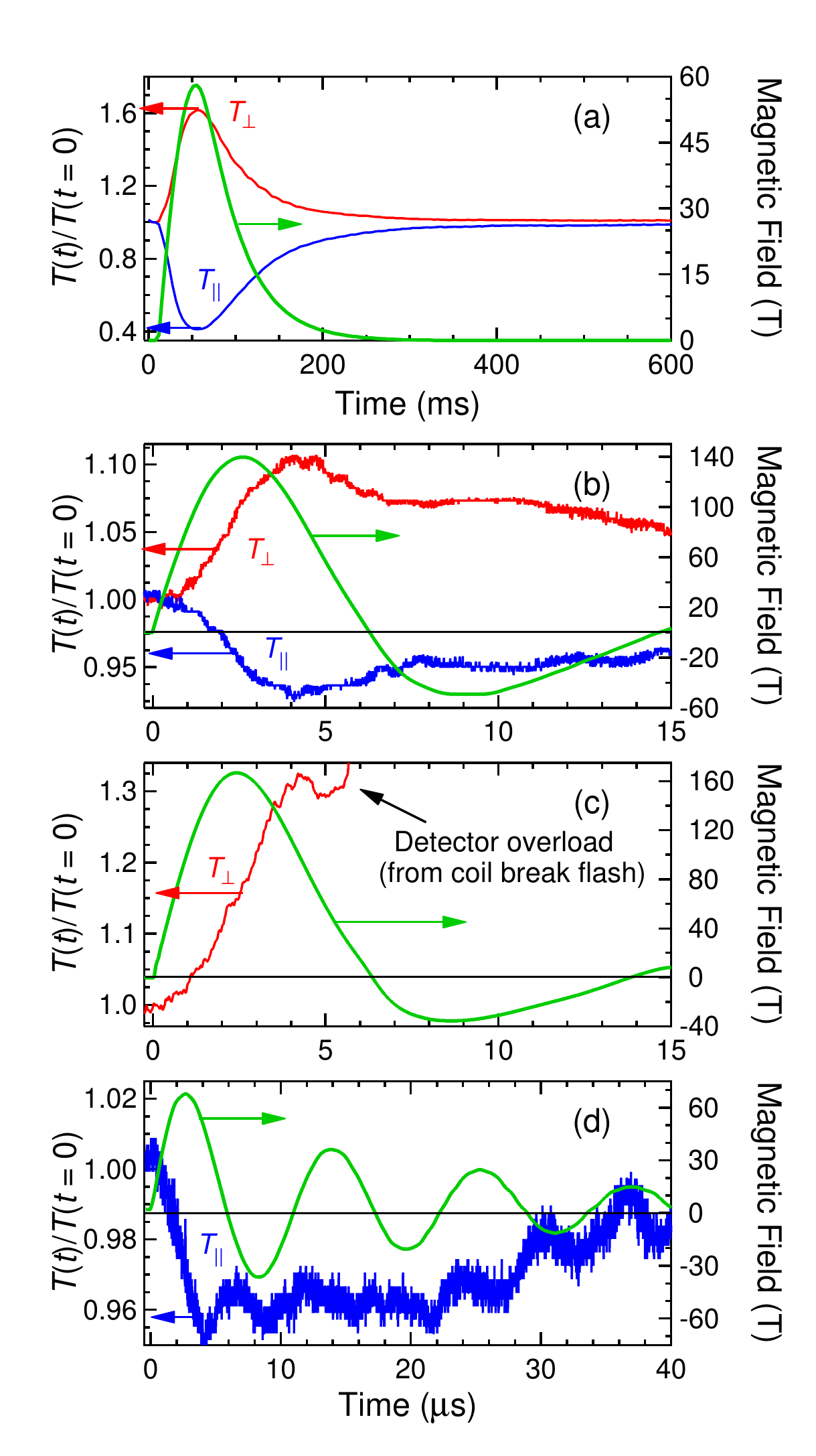}
\caption{(color online) Time-dependent traces of transmittance of light polarized parallel (red, left axis) and perpendicular (blue, left axis) to the applied magnetic field (green, right axis) for (a) a 56~T, 50-ms-rise-time pulse, (b) a 140~T, 2.5-$\mu$s-rise-time pulse (Megagauss), (c) a 166~T, 2.5-$\mu$s-rise-time pulse (STP) in the perpendicular polarization geometry, and (d) 65~T oscillating $\mu$s-field pulse and transmittance in parallel polarization geometry.  At zero magnetic field, the transmittances are equal.  As the field strength grows, the SWNTs align and decrease (parallel) or increase (perpendicular) the intensity of transmitted light.}
\label{transmittance}
\end{figure}

Figure~\ref{transmittance}(a) displays spectrally-integrated, time-dependent transmittance through the sample and polarizer [in parallel (blue) and perpendicular (red) configurations] and the accompanying 56~T magnetic field trace (green).  The raw transmittance data is normalized to the zero-field value as
\begin{equation}
T^N_{\parallel,\; \perp}(t) = \frac{T_{\parallel ,\; \perp} (t)}{T(t=0)}
\label{transmittance_eq}
\end{equation}
where $T_{\parallel,\; \perp}(t)$ denotes the raw transmittance as a function of time with the respective polarization configuration. Starting at time zero, before the field pulse, the transmittance in both polarization configurations is equal.  As the field increases, and the SWNTs start to align with the field, light polarized parallel (perpendicular) to the magnetic field decreases (increases) in overall transmittance.

Similarly, Fig.~\ref{transmittance}(b) shows the optical response of suspended SWNTs to a $\mu$s-pulse magnetic field produced by the Megagauss Generator in Berlin\cite{STCBerlin}.  This data was collected with an Ar$^+$ ion laser at 488~nm, which is in the second subband region of the SWNT optical spectra, and thus the Aharonov-Bohm-effect-induced spectral changes are small in relation to the linewidth, negating the need for spectral integration.  As the field rises to 140~T ($\sim$2.5~$\mu$s rise time), the nanotubes align to their maximum value, which lags the peak field by $\sim$2~$\mu$s.  It should be noted that in this experiment the field returns to zero at $\sim$~6~$\mu$s and then increases {\em in the negative direction}, reaching a minimum of $\sim-$50~T at $\sim$~9~$\mu$s.   However, since only the magnitude of the magnetic field ($|\vec{B}|$) is important in aligning the nanotubes, the transmittance shows a secondary peak at $\sim$10~$\mu$s. This is also clearly demonstrated by the parallel configuration data in Fig.~\ref{transmittance}(d) where we used the Megagauss Generator to produce a rapidly oscillating field of $\approx$65~T.  Figure~\ref{transmittance}(c) shows results from the Los Alamos Single Turn Coil Project (STP) magnet\cite{MielkeMcDonaldMG2006} in a perpendicular configuration with a 635~nm laser and a different sample. At approximately 6~$\mu$s,  when magnitude of the field was low, in part (c) the detector overloaded due to the arc flash from the routine disintegration of the coil, this does not affect the data collected before the coil break.  This data confirms our results from the Megagauss Generator with a different magnet of similar design, different excitation wavelength, and different sample.  Overall, the magnitude of the change in transmittance is less than the ms pulse experiment due to the shorter field duration.  Figures~\ref{transmittance}(a) and \ref{transmittance}(d) are nearly the same magnitude, but the $\mu$s-pulse in \ref{transmittance}(d) shows an order of magnitude smaller response than the ms-pulse in \ref{transmittance}(a).  For our qualitative analysis we use ms-pulse data from Toulouse and $\mu$s-pulse data from Berlin.
\afterpage{\clearpage}

\subsection*{Calculated Dynamic Linear Dichroism}

The time-dependent (or dynamic) linear dichroism, $LD(t)$, of SWNT alignment is calculated directly from the normalized transmittances.  Using the relationship between transmittance ($T$) and absorbance ($A$), $LD(t)$ can be related to the measured transmittances, $T_\parallel(t)$ and $T_\perp(t)$, as
\begin{equation}
 \begin{split}
 LD(t) &= A_\parallel(t) - A_\perp(t) \\
 &= -\ln{\frac{T_\parallel(t)}{T_0}} +\ln{\frac{T_\perp(t)}{T_0}} \\
 &= \ln{\frac{T_\perp(t)}{T_\parallel(t)}}
 \label{LDequation}
 \end{split}
\end{equation}
where the transmittance of the background medium, $T_0$, cancels out.  This is of particular advantage in pulsed field experiments, where the induced change in transmittance is very straightforward to collect, but the background signal can be cumbersome.  As we are studying the dynamics of SWNT alignment in pulsed fields, and not the magnitude of alignment, we can utilize $LD(t)$ normalized to its maximum value ($\overline{LD}(t) \equiv LD(t) / LD_{\rm max}$).  Although this procedure washes out the quantitative measure of the alignment as opposed to normalizing by isotropic absorption as in $LD^r=3S$, it retains the dynamics of the SWNTs in response to the magnetic field pulse.

\begin{figure}[!h]
\includegraphics[scale=.7]{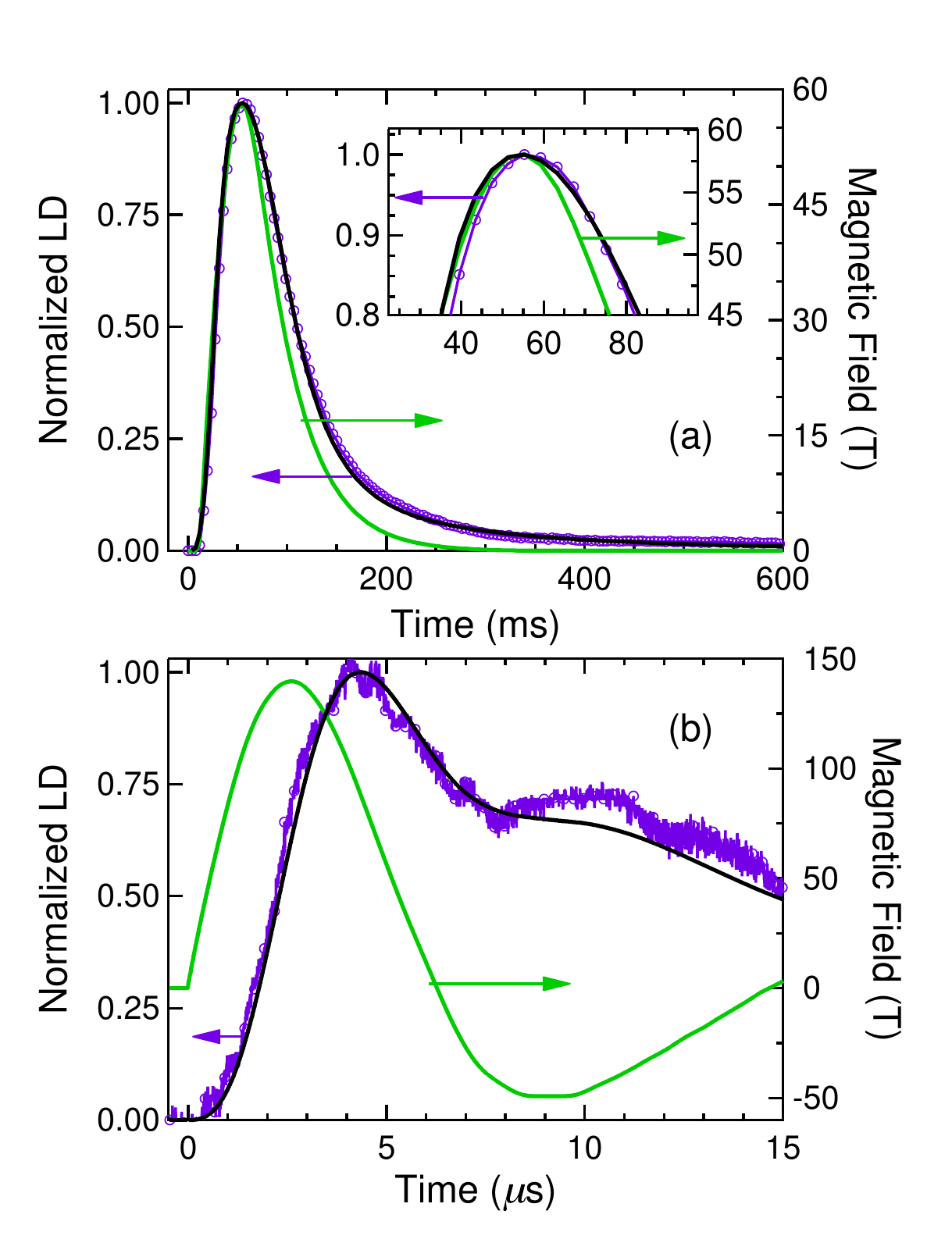}
\caption{(color online) Time-dependent traces of calculated normalized dynamic linear dichroism $\overline{LD}(t)$ (purple, left axis) and applied magnetic magnetic field (green, right axis) for (a) a 56~T, ms-pulse and (b) a 140~T, $\mu$s-pulse.  As the sample is isotropic at zero magnetic field, the linear dichroism is zero.  As the field strength grows in time and the SWNTs align with the magnetic field, the dichroism increases, peaking at a time slightly lagged to the maximum of the magnetic field.  After the magnetic field pulse, the sample gradually relaxes to its unaligned state.  Comparison to normalized linear dichroism computed from our model is shown in solid black.}
\label{LDfigure}
\end{figure}
Figure~\ref{LDfigure} shows $\overline{LD}(t)$ (purple) for (a) ms and (b) $\mu$s pulses calculated from the transmittances of Fig.~\ref{transmittance}.  The relationship of $\overline{LD}$ and $LD^r$ is such that they share the same dynamic features.  The positive sign of the signal indicates that the SWNTs are aligning with the magnetic field.  As the magnetic field increases to a strength greater than the randomization of the Brownian potential, the SWNTs feel a strong force to align.  However, there is a lag due to viscous drag, thus they always have a torque to align to the direction of the applied magnetic field.  As the magnetic field decreases, there is a point where the tubes will no longer increase in alignment (the point of maximum $LD$).  As the field decreases further, and the Brownian term becomes more significant, eventually the SWNTs randomize, slowed by viscous drag.  When the magnetic field is back to zero, starting from any residual alignment present in the sample, there is a competition between Brownian motion and the viscosity of the solution; this gives the characteristic relaxation time of the SWNTs.

\section*{Theory}
\begin{figure}[!h]
\includegraphics[scale=.75]{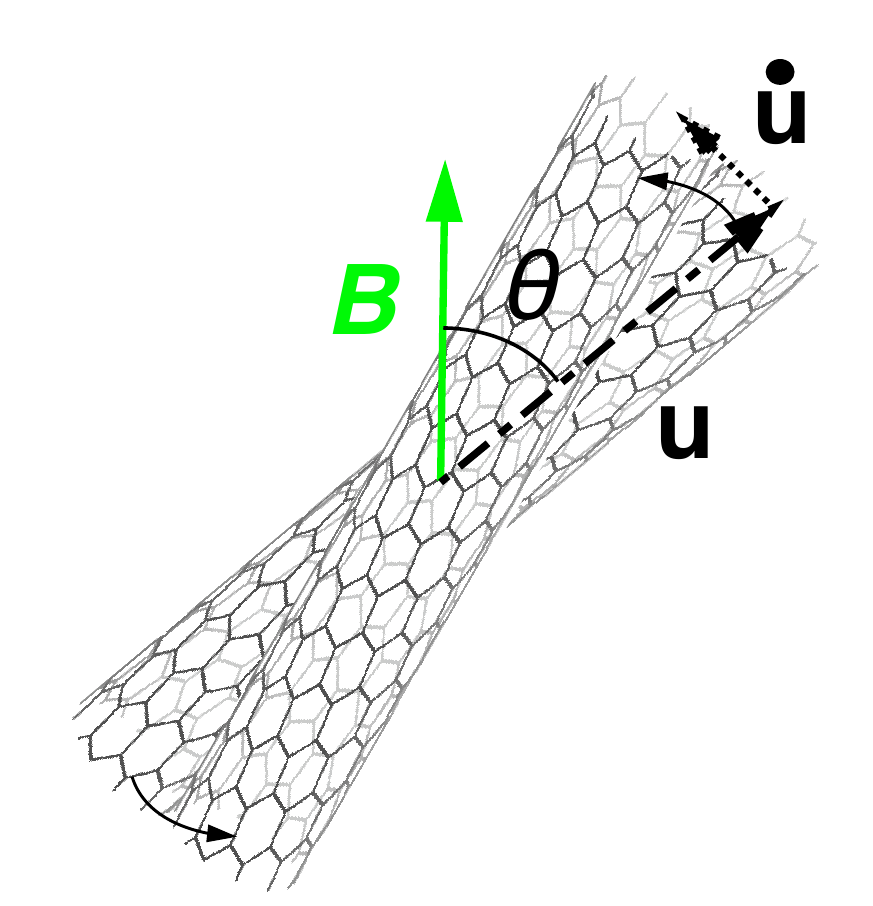}
\caption{A SWNT with a direction defined by the vector $\mathbf{u}$ at an angle $\theta$ to the magnetic field $B$; the magnetic properties of the SWNT creates a torque $N_{mag}$ forcing the SWNT to align with the magnetic field. Its direction changes as $\dot{\mathbf{u}}$, defining an angular velocity $\omega~=~\mathbf{u}~\times~\dot{\mathbf{u}}$}
\label{fig:u}
\end{figure}
In order to understand the effect of the magnetic field on the overall alignment of the SWNTs in solution, we must understand the competition between thermal agitation, or Brownian motion, which functions to randomize the nanotube orientation and the magnetic field, which functions to align the nanotubes.  Since the persistence length of a single SWNT is much greater than the length of the SWNTs in our study,\cite{duggal:246104} we consider SWNTs to behave like rigid-rods in suspension of radius $R$ and poly-disperse length $L$.  We examine a dilute dispersion of non-interacting SWNTs, which enables us to consider the orientation of each nanotube independently and determine the bulk orientation by summing the contributions from each nanotube in the distribution.

Figure~\ref{fig:u} depicts a SWNT oriented in the direction $\mathbf{u}$ at an angle $\theta$ to the magnetic field $B$; the SWNT orientation is dependent on the total torque,
\begin{equation}
N_{\rm tot} = N_{\rm Brown} + N_{\rm mag},
\end{equation}
which is the sum of contributions from Brownian motion and the magnetic field.  If $\Psi(L;\mathbf{u};t)$ is the probability distribution function of $\mathbf{u}$ and $U(L;\mathbf{u};t)$ is the external potential, then the Brownian motion contribution is included by adding $k_{\rm B} T \ln{\Psi}$ to $U$.  The angular velocity $\omega$ induced by the total torque is\cite{DoiEd1986}
\begin{equation}
\omega = \frac{1}{\varsigma_r} N_{\rm tot} = -\frac{1}{\varsigma_r}(k_BT\Re\ln{\Psi} + \Re U),
\label{eqn:w}
\end{equation}
where the rotational operator $\Re$ is defined as
\begin{equation}
\label{eqn:R}
\Re \equiv \mathbf{u}\times\frac{\partial}{\partial \mathbf{u}}
\end{equation}
and the rotational friction constant $\varsigma_r$ is defined as\cite{Larson1999}
\begin{equation}
\label{eqn:vr}
\varsigma_r = \frac{\pi \eta_s L^3}{3}\epsilon f(\epsilon)~,
\end{equation}
where
\begin{equation}
\epsilon = \left( \ln{\frac{L}{R}} \right)^{-1}
\label{eqn:feps}
\end{equation}
and
\begin{equation}
f(\epsilon) = \frac{1+0.64\epsilon}{1-1.5\epsilon} + 1.659\epsilon^2~.
\end{equation}
The equation for the conservation of the probability distribution $\Psi$ then becomes
\begin{equation}
\label{eqn:smo} 
\frac{\partial \Psi}{\partial t} = -\Re \cdot (\omega\Psi) = D_r\Re\cdot [\Re\Psi + \frac{\Psi}{k_BT}\Re U]~,
\end{equation}
where the rotational diffusion is defined as
\begin{equation}
\label{eqn:dr}
D_r  = \frac{k_BT}{\varsigma_r}.
\end{equation}
Eq.~(\ref{eqn:smo}) is known as the Smoluchowski equation for rotational diffusion\cite{DoiEd1986}.

In our system the external potential is the magnetic field's effect on the orientation of an individual SWNT.  This potential depends on the magnetic susceptibility anisotropy, $\Delta\chi$, of the SWNT, the number of carbon atoms in the SWNT, $N(L)$, the strength of the magnetic field, $B(t)$, and the orientation of the SWNT as measured by the angle $\theta(\mathbf{u})$:
\begin{equation}
\label{eqn:U}
U(L;\mathbf{u};t) = -\Delta\chi N(L) B(t)^2 \cos^2{\theta(\mathbf{u})}.
\end{equation}

To track the nematic order parameter $S(t)$ of the SWNT suspension in a time-dependent magnetic field, we first solve the Smoluchowski equation by expanding $\Psi$ as a sum of spherical harmonics $Y^m_n$:
\begin{equation}
\label{eqn:psi}
\Psi(L;\mathbf{u};t) = \sum_{n=0,2}^{N} \sum_{m=-n,2}^{n} A^m_n(L;t) Y^m_n(\mathbf{u}).
\end{equation}
Spherical harmonics are ideal basis functions because they are eigenfunctions of the highest derivative operator in Eq.~(\ref{eqn:smo}). Note that only the even values of $n$ are used because the system is symmetric about the alignment axis.  Note also that only the even values of $m$ are needed since the SWNTs have no permanent magnetic moments (they have only induced magnetic dipoles), and so $\Psi(\mathbf{u}) = \Psi(-\mathbf{u})$.\cite{StewartetAl72JRa}  

The energy can be expressed simply in terms of the second spherical harmonic $Y^0_2$
\begin{eqnarray}
U=\Delta\chi N B^2 \cos^2{\theta} & = & \Delta\chi N B^2 \left[\frac{4}{3}\sqrt{\frac{\pi}{5}} \left(Y_2^0+\frac{1}{3}\right)\right] \nonumber \\
& = & \kappa \left(Y_2^0+\frac{1}{3}\right)
\label{eqn:UY2}
\end{eqnarray}
where
\begin{equation}
\kappa(L;t) = \frac{4}{3}\sqrt{\frac{\pi}{5}} \Delta\chi N(L) B(t)^2
\label{eqn:alpha}
\end{equation}

The partial differential equations, Eq.~(\ref{eqn:smo}), are then converted into a system of ordinary differential equations for $A^m_n$ using Galerkin's method. By multiplying Eq.~(\ref{eqn:smo}) by each basis function $Y^p_q$ and integrating over all space, the time evolution of each corresponding coefficient, $\frac{d}{dt}A_q^p$, can be determined as
\begin{eqnarray}
\label{eqn:gal}
\int \sin{\theta} d \theta \int d \phi \; Y_q^p \; \frac{d\Psi}{dt} = \int \sin{\theta} d \theta \int d \phi \; Y_q^p \; \frac{d}{dt} \sum_{n=0,2}^{N} \; \sum_{m=-n,2}^{n} A_n^m Y_n^m = \frac{d}{dt}A_q^p \nonumber \\
~\\ 
=-D_rq(q_1)A_q^p-6\kappa \frac{D_r}{k_BT} \sum_{n=0,2}^{N} \; \sum_{m=-n,2}^{n} A_n^m \int \sin{\theta} d \theta \int d \phi \; Y_p^q \; Y_n^m \; Y_2^0 \nonumber \\
-\kappa \frac{D_r}{k_BT}\sum_{n=2,2}^{N} \; \sum_{m=-n+2,2}^{n} A_n^m \sqrt{\frac{3}{2}(n-m)(n+m+1)} \int \sin{\theta} d \theta \int d \phi \; Y_p^q \; Y_n^{m-1} \; Y_2^1 \nonumber \\
-\kappa \frac{D_r}{k_BT}\sum_{n=2,2}^{N} \; \sum_{m=-n,2}^{n-2} A_n^m \sqrt{\frac{3}{2}(n+m)(n-m+1)} \int \sin{\theta} d \theta \int d \phi \; Y_p^q \; Y_n^{m+1} \; Y_2^{-1}, \nonumber \\
\end{eqnarray}
where the integrals of the multiplication of three spherical harmonics, i.e., $\int \sin{\theta} d \theta \int d \phi \; Y_p^q \; Y_n^{m+1,m,m-1} \; Y_2^{-1,0,1}$, are nonzero only when $m=p=0$ or $p=-m$.\cite{Arfken1985}  The initial values of the coefficients are determined from the initial orientation of the nanotubes; a random orientation is described by $A_n^m=0$ except for $A_0^0=1$.  The magnetic field is turned on at $t=0$ and varies with time.  The coefficients at each time step are solved by using a numerical ordinary differential equation integration technique -- third-order Runge-Kutta, available in MATLAB (ODE23).  $S$ is related with the coefficients, $A_n^m(L)$, by averaging over $\cos^2{\theta(L)}$,
\begin{eqnarray} 
& \ & \left<\cos^2{\theta(L)}\right> = \left< \frac{4}{3}\sqrt{\frac{\pi}{5}} Y_2^0+\frac{1}{3}\right> \nonumber \\ 
& = & \int\int \left(\frac{4}{3}\sqrt{\frac{\pi}{5}} Y_2^0+\frac{1}{3}\right) \sum_{n=0,2}^{N} \; \sum_{m=-n,2}^{n} A_n^m \Psi_n^m \sin{\theta} \; d\theta \; d\phi \nonumber \\ 
& = & \frac{4}{3}\sqrt{\frac{\pi}{5}} A_2^0 \int\int  (Y_2^0)^2 \sin{\theta} \; d\theta \; d\phi +  \frac{1}{3} A_0^0 \int\int\sin{\theta} \; d\theta \; d\phi  \nonumber \\ 
& = & \frac{4}{3}\sqrt{\frac{\pi}{5}} A_2^0 +\frac{1}{3}. \label{eqn:cos}
\end{eqnarray}
By placing Eq.~(\ref{eqn:cos}) into Eq.~(\ref{orderparam}), we find $S(L,t)$ to be
\begin{equation}
\label{eqn:S}
S(L;t) = 2\sqrt{\frac{\pi}{5}} A_2^0(L;t).
\end{equation}

The bulk solution's nematic order parameter $S(t)$ is determined by integrating $S(L;t)$ over the distribution of lengths
\begin{equation}
\label{eqn:SL}
S(t) = 2\sqrt{\frac{\pi}{5}} \int_0^\infty A_2^0(L,t) \Omega(L) dL.
\end{equation}
To compare with experimental data, we assume a lognormal probability distribution,
\begin{equation}
\label{eqn:dist}
\Omega(L) = \frac{1}{L\sigma\sqrt{2\pi}}e^{-\frac{(\ln{L}-\mu)^2}{2\sigma^2}},
\end{equation}
and vary the parameters $\mu$ and $\sigma$, the mean and standard deviation of $\log{L}$, respectively, to calculate $\overline{LD}(t)$ = max$(S(t))/S(t)$, which is compared with our measured $\overline{LD}(t)$.

\section*{Discussion}
\begin{figure}[!h]
\includegraphics[scale=.7]{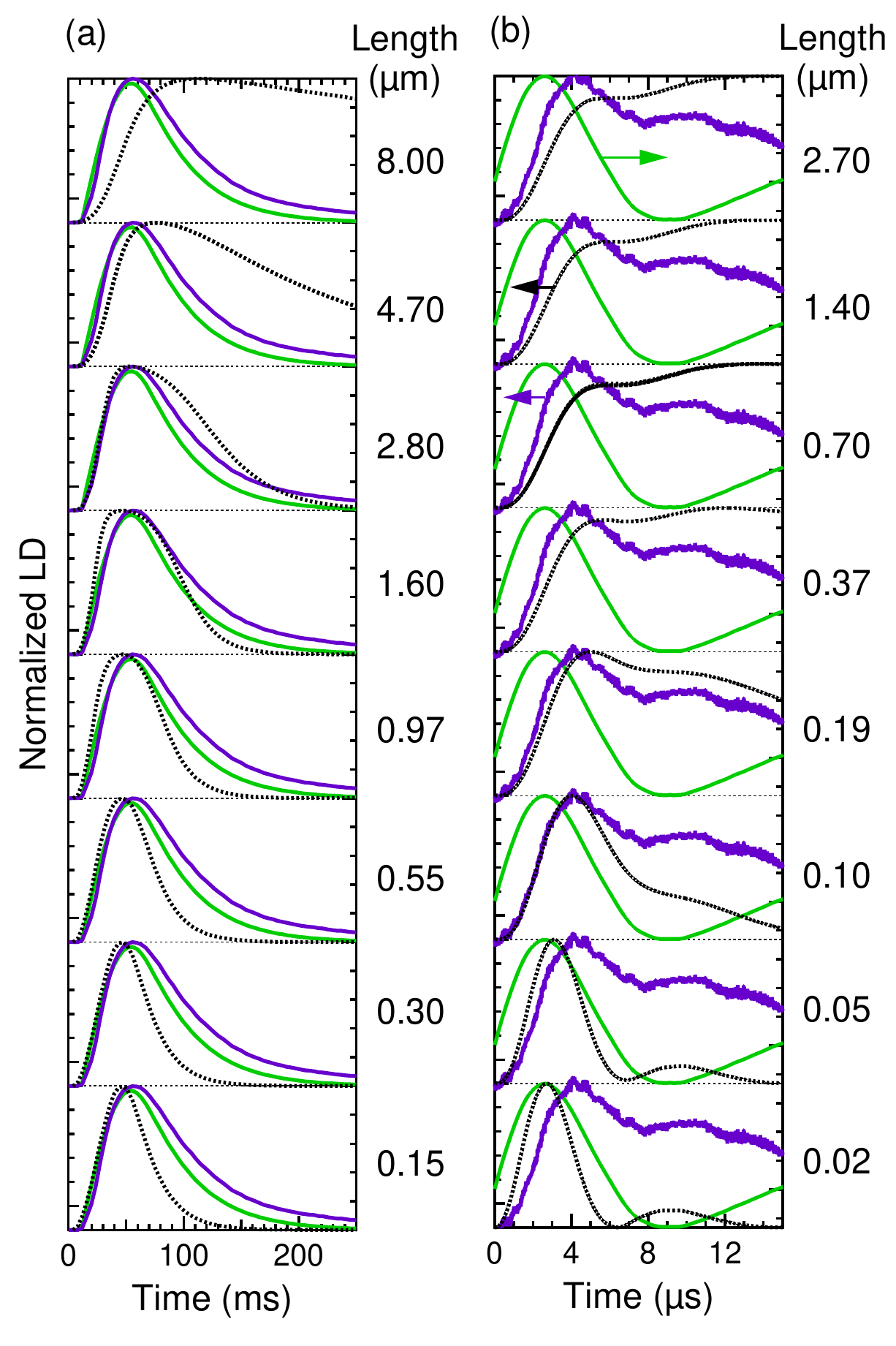}
\caption{(color online) The contributions from each length in the distribution normalized to their maximum value (dotted black), the experimental $\overline{LD}$ (purple), and the accompanying magnetic field pulses (green).  Traces are offset for clarity, and the field and experimental traces are reproduced at each offset for ease of comparison.  Note that no single dotted trace can successfully reproduce the experimental $\overline{LD}$.  Part (a): ms-pulse, part (b): $\mu$s-pulse.}
\label{lengthcontributions_norm}
\end{figure}

We can now use our model to calculate the dynamic response of SWNTs in time-varying magnetic fields and compare with the experimental data.  Figure~\ref{lengthcontributions_norm} compiles simulated $\overline{LD}$ for several lengths.  Each simulated $\overline{LD}$ trace (dotted black) is offset vertically and plotted along side its applied magnetic field (green) and experimental $\overline{LD}$ (purple).  In general, shorter nanotubes have less viscous drag, and hence, align to the field pulse faster, but also randomize faster as they have less $\Delta\chi$.  Longer nanotubes take longer to respond to the field, as they have more viscous drag in solution, but their overall alignment is larger due to their larger $\Delta\chi$.  These effects are also convolved with the duration and strength of the field impulse.  A shorter impulse will more readily align short tubes than long tubes during the pulse duration.  Figure~\ref{lengthcontributions_norm}(a) shows the ms-pulse data while Fig.~\ref{lengthcontributions_norm}(b) displays the $\mu$s-pulse.

\begin{figure}[!h]
\includegraphics[scale=.75]{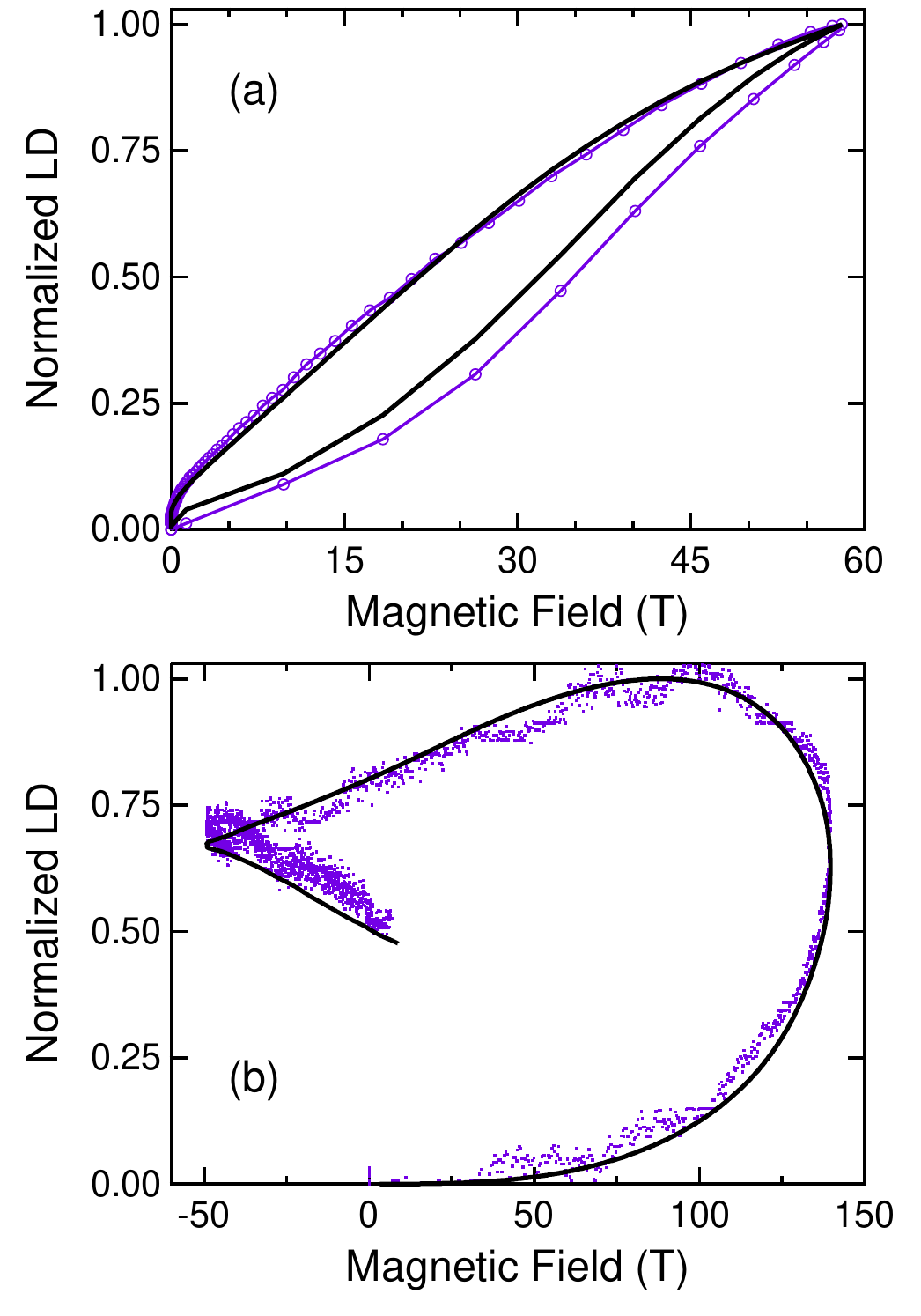}
\caption{(color online) Magnetic field dependent traces of calculated (purple) and simulated (black) normalized linear dichroism vs. applied magnetic field.  The hysteresis is indicative of the lag to the magnetic field produced by our poly-disperse length sample.  Part (a) shows a 56~T, ms-pulse and (b) shows a 140~T, $\mu$s-pulse.}
\label{NDLD_B}
\end{figure}

\begin{figure}[!h]
\includegraphics[scale=.73]{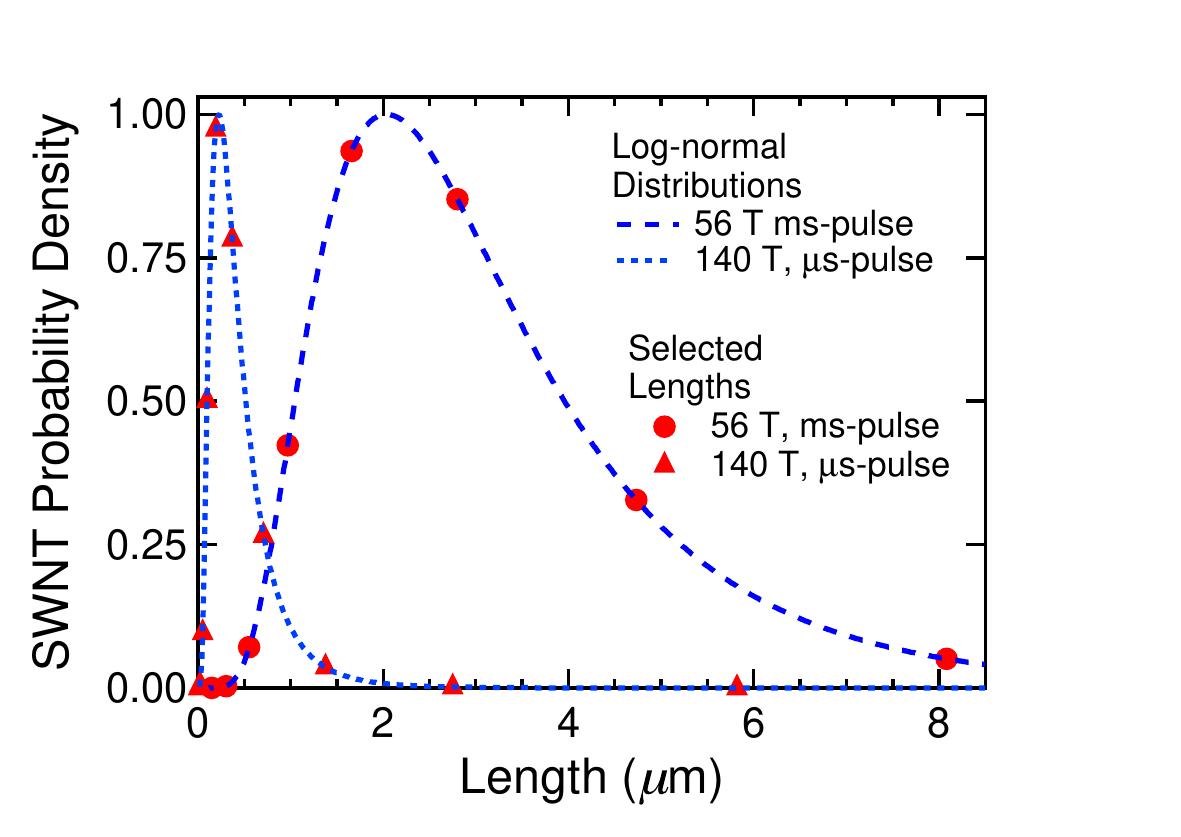}
\caption{(color online) Histograms of log-normal length distributions used to compute the simulated linear dichroism for each magnetic field pulse. The contributions from selected lengths in the distribution are noted by filled circles and triangles.}
\label{lengthdistributions}
\end{figure}

Due to the fact we have a sample that is polydisperse in length, as expected, no individual simulated length is able to reproduce all the features of the experimental data, as shown in Fig.~\ref{lengthcontributions_norm}.  To describe a typical SWNT length distribution, we use a log-normal form, which has been measured and confirmed by AFM and rheology measurements on similarly prepared samples.\cite{ParraVasquez:2007p162}  In Fig.~\ref{lengthdistributions} the lengths indicated by symbols are those that were explicitly calculated to determine the overall $\overline{LD}$ that best fit our experiment.  Figure~\ref{NDLD_B} compares the experimental $\overline{LD}$ signal with that obtained from our simulation as a function of magnetic field.  Our model shows a good overall match to the measured data using published values\cite{ZaricetAl04NL} for $\Delta\chi$, the corresponding alignment potential from Eq.~(\ref{eqn:U}), and the length distribution, $\Omega(L)$ from Eq.~(\ref{eqn:dist}). These results were obtained by varying average, $\mu$, and standard deviation, $\sigma$, of the natural log of $L$ in a log-normal distribution (Fig.~\ref{lengthdistributions}).

The comparisons in Fig.~\ref{LDfigure} are fit by the length distributions of Fig.~\ref{lengthdistributions}.  Figure~\ref{lengthcontributions_norm} gives an indication of which population of SWNTs is responsible for each part of the simulated $\overline{LD}$.  Shorter nanotubes are the predominant source of signal during the upsweep of the field and longer nanotubes for the down sweep (and lag).  As the samples were not from the same batch for the different time duration pulses, a rigorous comparison between these effects cannot be made.  Nonetheless, it is feasible to conclude that a shorter duration pulse will be moving predominantly individual nanotubes as our fit length distribution~\cite{ParraVasquez:2007p162} is close to published values.  The $\mu$s-pulse experiment is of too short duration to appreciably align very long SWNTs, so it is not sensitive to possible bundles in solution.  The ms-pulse experiment on the other hand is long enough to move large nanotubes but shows a slight mismatch on the upsweep of the magnetic field (Fig.~\ref{NDLD_B}).  It is possible that a bi-modal length distribution exists in solution, a population of shorter individualized nanotubes and one of longer bundles of nanotubes.  Further experiments on samples of known length distribution, measuring $LD^r$, are needed to investigate this hypothesis. 

\section*{Conclusion}
We have measured the magnetic-field-induced dynamic linear dichroism of SWNT solutions.  Our presented technique establishes a method for the extraction of the length distribution of the SWNTs present in solution based on the Smoluchowski equation.  However, future work is needed, specifically comparison with other techniques for determining length distributions, such as rheology and AFM measurements, allowing for refinement of published values of SWNT magnetic susceptibility and chirality dependence.  It is also possible from this work to design experiments that will predominantly probe certain lengths of SWNTs in solution, and investigate the possibility of varying length distributions with chirality. 

This work was supported by the Robert A.~Welch Foundation (through grant Nos.~C-1509 and C-1668), the National Science Foundation (through Grants Nos.~DMR-0134058, DMR-0325474, OISE-0437342, CTS-0134389, and CBET-0508498), and EuromagNET (EU contract RII3-CT-2004-506239).  We thank the support staff of the Rice Machine Shop, Institut f\"ur Physik, NHMFL, and LNCMP.  We also thank Scott Crooker and Erik Hobbie for helpful discussions.

\section*{\normalsize Methods}
\small{
HiPco SWNTs were suspended in aqueous surfactant solutions of sodium dodecylbenzene sulfonate (SDBS) using standard techniques.\cite{OconnelletAl02Science}  It is noted that the ultracentrifugation step in our preparation procedure minimizes the presence of ferromagnetic catalyst particles, which have been shown to have a strong effect on SWNT alignment in low DC magnetic field fields.\cite{islam:201401}  Samples were loaded into home-built cuvettes with path lengths of $\sim$1 to 2~mm before being inserted into one of the experimental transmittance setups used.

Short pulse magnetic field (56~T, ms-pulse) data was obtained at the Laboratoire National des Champs Magn\'{e}tiques Puls\'{e}s in Toulouse, France.  A broad band, quartz tungsten halogen (QTH) lamp was used with a fiber-coupled, Voigt geometry, transmittance probe with an adjustable polarizer.  Light transmitted through the polarizer (either parallel or perpendicular to the magnetic field) and sample was dispersed on a fiber-coupled 300-mm monochromator and detected with a liquid-nitrogen-cooled InGaAs diode array with a typical exposure time of $\sim$1~ms.  The magnetic field was generated by a $\sim$150~ms current pulse, using $\sim24\%$ of the energy from a 14~MJ capacitor bank, into a $\sim$26~mm free bore reinforced copper coil cooled to liquid nitrogen temperature, designed for 60~T pulses.  As the coil was at liquid nitrogen temperature before each experiment, a cryostat was utilized to keep the samples maintained at room temperature.

Megagauss measurements ($\mu$s-pulse) were performed at two installations: the Megagauss Generator\cite{megagauss_note,STCBerlin} ($\sim$140~T) at Humboldt-Universit\"at zu Berlin and the Single Turn Coil Project (STP) magnet\cite{STP_note} ($\sim$166~T) at the National High Magnetic Field Laboratory (NHMFL) in Los Alamos.  The Megagauss Generator and the STP magnet are single-turn coil magnets of similar design. They each utilize low inductance capacitor banks ($\sim$225~kJ in Berlin and 259~kJ in Los Alamos) capable of discharging $\sim$3.8~MA on a $\mu$s time-scale through a 15~mm or 10~mm single-turn copper coil.  These experiments are deemed ``semi-destructive,'' as the massive amount of current and huge Lorentz force on the conductor causes an outward expansion followed by explosion of the coil, ideally preserving the sample and sample holder for repeated use.  Oscillating fields were realized by preventing coil expansion through reinforcement.  Since the duration of the field in megagauss experiments was $\approx10^{-4}$ that of a long-pulse experiment, transmittance data was collected with higher intensity, single wavelength lasers.  An Ar$^+$ ion laser at 488~nm was utilized in Berlin and a diode laser at 635~nm was used in Los Alamos.  Light transmitted through a fiber coupled sample holder, cuvette, and polarizer, with similar geometries to the long pulse experiment, was collected on a Si photodiode (3~ns rise-time) connected to a fast oscilloscope using the sophisticated setup of reference~\cite{STCBerlinData}.  The measurements were done at room temperature, without the need of a cryostat.}


\providecommand{\refin}[1]{\\ \textbf{Referenced in:} #1}

\end{document}